\documentclass[aps,jmp,amsmath,amssymb,doublespace,reprint]{revtex4-1}

\usepackage{graphicx}
\usepackage{dcolumn}
\usepackage{bm}
\usepackage{lineno}
\usepackage{lipsum}
\usepackage[colorlinks=true,citecolor=blue]{hyperref}
\usepackage{resizegather}

\begin{document}

\title{Addition to ``Structure and dynamics of a polymer-nanoparticle composite: Effect of nanoparticle size and volume fraction"}

\author{Valerio Sorichetti}
\affiliation{Laboratoire Charles Coulomb (L2C), Univ. Montpellier, CNRS, F-34095, Montpellier, France and IATE, INRA, CIRAD, Montpellier SupAgro, Univ. Montpellier, F-34060, Montpellier, France}
\author{Virginie Hugouvieux}
\affiliation{IATE, INRA, CIRAD, Montpellier SupAgro, Univ. Montpellier, F-34060, Montpellier, France}
\author{Walter Kob}
\affiliation{Laboratoire Charles Coulomb (L2C), Univ. Montpellier, CNRS, F-34095, Montpellier, France}

\date{\today}

\begin{abstract}
In our previous publication (Ref.~\citenum{sorichetti2018structure}) we have shown that the data for the normalized diffusion coefficient of the polymers, $D_p/D_{p0}$, falls on a master curve when plotted as a function of $h/\lambda_d$, where $h$ is the mean interparticle distance and $\lambda_d$ is a dynamic length scale. In the present note we show that also the normalized diffusion coefficient of the nanoparticles, $D_N/D_{N0}$, collapses on a master curve when plotted as a function of $h/R_h$, where $R_h$ is the hydrodynamic radius of the nanoparticles.
\end{abstract}
\maketitle

\begin{figure}
\centering
\includegraphics[width=0.45 \textwidth]{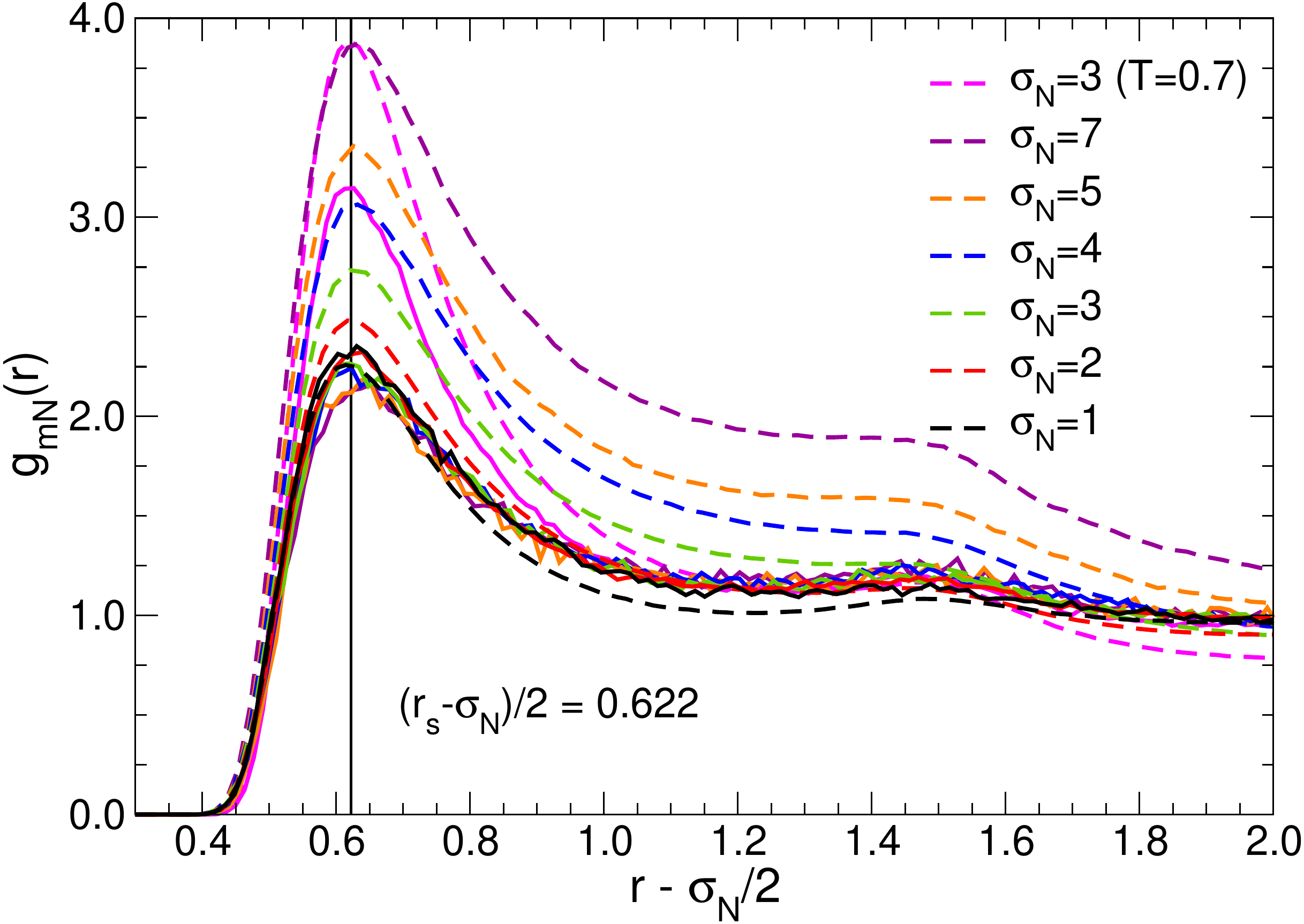}
\caption{Monomer-NP radial distribution function $g_{mN}(r)$ for different values of the nanoparticle diameter $\sigma_N$, temperature $T$, and nanoparticle volume fraction $\phi_N$.  For each pair of $\sigma_N,$ and $T$ the dashed curves correspond to the highest values of $\phi_N$ considered in Ref.~\citenum{sorichetti2018structure}, whereas the continuous curves correspond to the lowest values (see Tab.~S2 in the Supporting Information of Ref.~\citenum{sorichetti2018structure} for the precise values for $T=1.0$). Curves of the same color have the same $\sigma_N$ and $T$. All the curves are for $T=1.0$ except where explicitly stated.}
\label{rdf_pm}
\end{figure}

In Sec.~4.4 of the original paper \cite{sorichetti2018structure} we discussed the behavior of the reduced diffusion coefficient of the nanoparticles (NP), $D_N/D_{N0}$, as a function of the NP volume fraction $\phi_N$ (Fig.~13a) and of the NP diameter $\sigma_N$ (Fig.~13b). Here, $D_{N0}=\lim_{\phi_N \to 0} D_N$.

We report here an interesting observation that went unnoticed before the publication of Ref.~\citenum{sorichetti2018structure}. In that paper, Fig.~11, we showed that the reduced diffusion coefficient of the polymers $D_p/D_{p0}$ collapses on a master curve when plotted as a function of the dynamic confinement parameter $h/\lambda_d$, where $h$ is the interparticle distance (see Sec.~AI in the Appendix of Ref.~\citenum{sorichetti2018structure}) and $\lambda_d$ is a fit parameter that depends on temperature and which can be interpreted as a dynamic length scale associated to polymer motion. The parameter $h/\lambda_d$ is a generalization of the static confinement parameter introduced by Composto and coworkers \cite{gam2011macromolecular,lin2016macromolecule}. The master curve is well described by the expression $D_p/D_{p0}=1-\exp(-h/\lambda_d)$ (Eq.~14 in Ref.~\citenum{sorichetti2018structure}).

Applying the same line of reasoning to the NP diffusion coefficient, we have fitted $D_N/D_{N0}$ to the expression $1-\exp(-h/\lambda_N)$ and found that this expression fits the data well with $\lambda_N \simeq \sigma_N+1.2$. In the following, we will argue that this value can be interpreted as the hydrodynamic radius of the NPs, i.e., $\lambda_N = R_h$. 

In Ref.~\citenum{sorichetti2018structure}, Sec.~4.4, we defined the hydrodynamic radius of the NPs as $R_h=(\sigma_N+1)/2$, which is the minimum distance between the center of a NP and that of a monomer if they are considered as hard spheres of diameter $\sigma_N$ and $1$, respectively. However, the monomer-NP interaction potential is certainly steep, but not a hard-sphere potential. A more precise way to define $R_h$ is therefore to resort to the monomer-NP radial distribution function, $g_{mN}(r)$, which we present in Fig.~\ref{rdf_pm}. This graph shows that for all values of $\phi_N$, $\sigma_N$ and temperature $T$ considered in the original work, $g_{mN}(r)$ displays a main peak located at 

\begin{equation}
r=\frac{r_s} 2 = \frac {\sigma_N + (2^{7/6} -1)} 2 \simeq \frac{ \sigma_N+1.245} 2 = \frac{\sigma_N} 2 + 0.622,
\end{equation}

\noindent where $r_s$, defined in Sec.~3.1 of Ref.~\citenum{sorichetti2018structure}, is twice the distance of the minimum of the monomer-NP potential. The position of the main peak corresponds to the typical distance of closest approach between a monomer and a NP. Since Fig.~\ref{rdf_pm} shows that the position of the peak relative to $\sigma_N$ is basically independent of the relevant parameters of the system, i.e., $\phi_N$ and $T$, and since it has been obtained directly by analyzing the configurations (with no assumption on the monomer-NP interaction potential) we propose the identification $R_h=r_s/2$ \footnote{We note that this redefinition of $R_h$ does not change the results discussed in Sec.~4.3 of Ref.~\citenum{sorichetti2018structure} (notably Fig.~12) significantly, since the value of $\sigma_h/2R_g=R_h/R_g$ changes only slightly.}, thus 

\begin{equation}
    R_h=\frac{\sigma_N}{2}+0.622 .
    \label{eq:Rh}
\end{equation}

\begin{figure}[t]
\centering
\includegraphics[width=0.45 \textwidth]{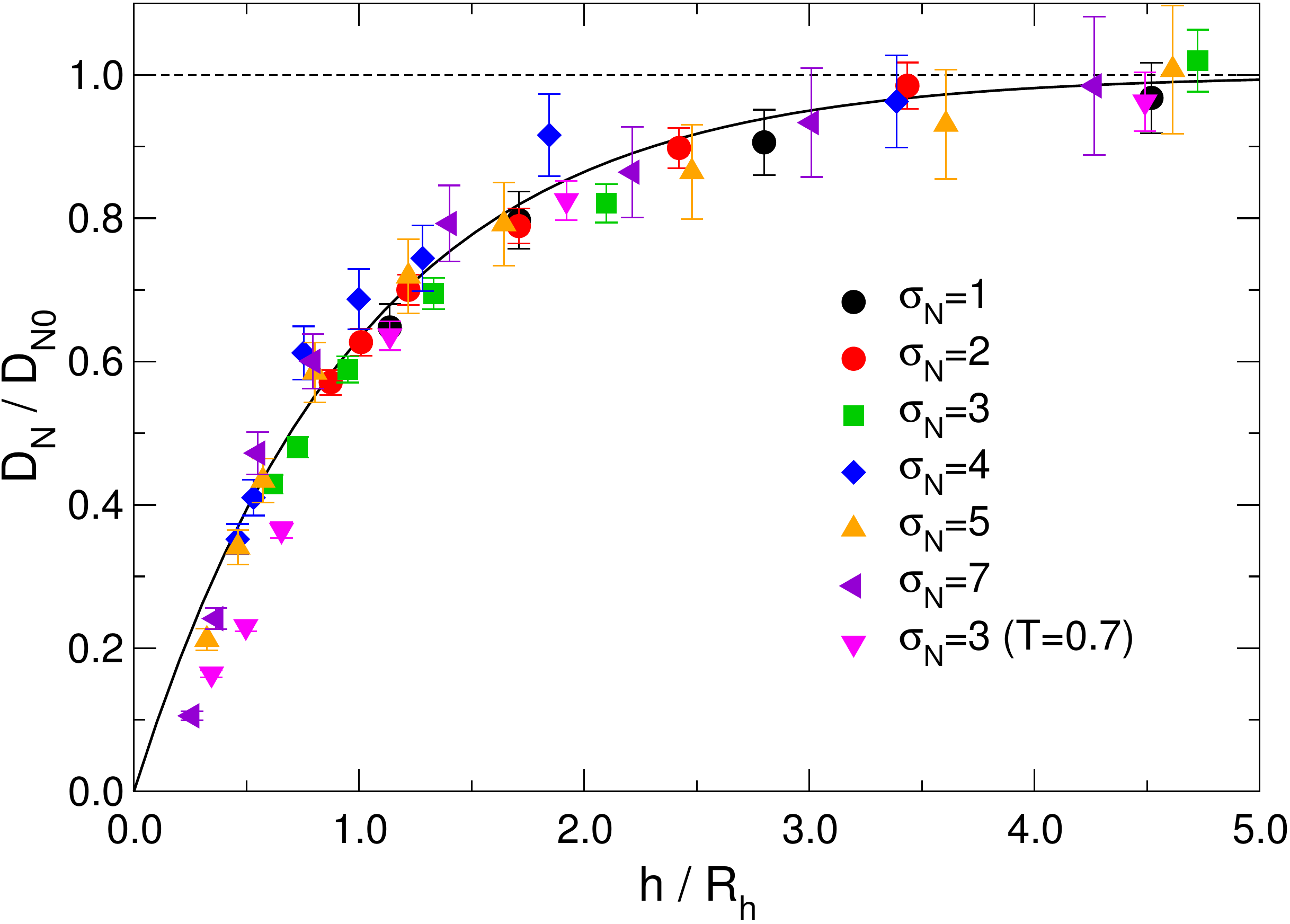}
\caption{Reduced NP diffusion coefficient $D_N/D_{N0}$ as a function of $h/R_h$, where $h$ is the interparticle distance and $R_h$ is the hydrodynamic radius of the NPs, Eq.~\eqref{eq:Rh}. Continuous line: Eq.~\eqref{dnp_curve}.}
\label{dnp_norm}
\end{figure}

In Fig.~\ref{dnp_norm} we show $D_N/D_{N0}$ as a function of $h/R_h$, and compare it with the expression

\begin{equation}
\frac{D_N}{D_{N0}} = 1-\exp\left(-\frac h {R_h}\right).
\label{dnp_curve}
\end{equation}

\noindent We stress that here $R_h$ is not a fit parameter but obtained from Eq.~(\ref{eq:Rh}). We note that although Eq.~\eqref{dnp_curve} gives a good description of the data, the discrepancy between the data and Eq.~\eqref{dnp_curve} is not negligible, as it is clear from the fact that one sees small but significant differences between the data and Eq.~\eqref{dnp_curve} at small values of $h/R_h$. 

In conclusion, we have shown that the interparticle distance $h$ controls not only the dynamics of the polymers, but also that of the NPs. In the case of the polymers, the reduced diffusion coefficient $D_p/D_{p0}$ collapses on a master curve when plotted as a function of $h/\lambda_d$, where $\lambda_d$ is a temperature-dependent parameter which does not depend on the NP diameter $\sigma_N$. In the case of the NPs, the variable which controls $D_N/D_{N0}$ is $h/R_h$, where $R_h$ is the hydrodynamic radius of the NPs, given by Eq.~(\ref{eq:Rh}), which we find to be basically independent of temperature. At present, it is not clear how general these results are and if there is a unifying  theory which explains both these observations, and therefore additional studies on this dependence are called for. In particular, simulations at different thermodynamic conditions and possibly with chains of different lengths are needed in order to clarify these questions.

\bibliography{a_main.bib}

\end{document}